\documentclass[aps,prb,twocolumn,showpacs,superscriptaddress,amsmath,amssymb,reprint]{revtex4-1}

\usepackage{graphicx}
\usepackage{epstopdf}
\usepackage{amsmath}
\usepackage{amsfonts}
\usepackage{amssymb}
\usepackage{bm}
\usepackage[utf8]{inputenc}
\usepackage{color}
\definecolor{blue(pigment)}{rgb}{0.15, 0.15, 0.7}
\def\gsim{\lower.35em\hbox{$\stackrel{\textstyle>}{\textstyle\sim}$}}
\def\lsim{\lower.35em\hbox{$\stackrel{\textstyle<}{\textstyle\sim}$}}

\usepackage{appendix}
\usepackage{url}
\usepackage[hyperfootnotes=false,hypertexnames=false]{hyperref}
\hypersetup
{
	pdftitle={Drift-induced modifications to the dynamical polarization of graphene},
	bookmarksnumbered=true,
	bookmarksopen=true,
	bookmarksopenlevel=1,
	colorlinks=true,
	citecolor=blue(pigment),
	linkcolor=blue(pigment),
	urlcolor=blue(pigment)
}

\begin{document}

\title{Drift-induced modifications to the dynamical polarization of graphene}

\author{Mohsen Sabbaghi}
\affiliation{Center for Superfunctional Materials, Department of Chemistry and Department of Physics, Ulsan National Institute of Science and Technology (UNIST), Ulsan 689-798, Korea}
\affiliation{Department of Physics, Pohang University of Science and Technology, Pohang 790-784, Korea}
\affiliation{Instituto de Ciencia de Materiales de Madrid, CSIC, E-28049 Madrid, Spain}

\author{Hyun-Woo Lee}
\email{hwl@postech.ac.kr}
\affiliation{Department of Physics, Pohang University of Science and Technology, Pohang 790-784, Korea}

\author{Tobias Stauber}
\email{tobias.stauber@csic.es}
\affiliation{Instituto de Ciencia de Materiales de Madrid, CSIC, E-28049 Madrid, Spain}

\author{Kwang S. Kim}
\email{kimks@unist.ac.kr}
\affiliation{Center for Superfunctional Materials, Department of Chemistry and Department of Physics, Ulsan National Institute of Science and Technology (UNIST), Ulsan 689-798, Korea}

\date{\today}

\begin{abstract}
The response function of graphene is calculated in the presence of a constant current across the sample. For small drift velocities and finite chemical potential, analytic expressions are obtained and consequences on the plasmonic excitations are discussed. For general drift velocities and zero chemical potential, numerical results are presented and a plasmon gain region is identified that is related to interband transitions. 
\end{abstract}


\maketitle
\section{Introduction}\label{sec:INTRO}
The prominent light-matter interaction of graphene has attracted immense interest due to the tunable plasmonic excitations in the THz \cite{Ju11,Fei11,Chen12,Fei12,Crassee12,Strait13} and mid-infrared \cite{Brar13,Woessner15} regimes. These charge density excitations in graphene, which are mostly explained by the density-density and current-current correlation functions of its $\pi$ electron gas \cite{STAUBER_10}, are well-explored in both theoretical \cite{FNDOPE,HWANG,PRINCIPI,MIKHAILOV} and experimental \cite{EXP_RES_1,EXP_RES_2,Liu} directions.

The manipulation of the density of states (DOS) of the $\pi$ and $\pi^{*}$ bands of graphene can be a tool for tailoring its plasmonic excitations\textemdash{}a scenario realizable through the exposure of graphene to either mechanical stress \cite{STRDOS_1,STRDOS_2} or perpendicular magnetic field \cite{MAGDOS_1,MAGDOS_2,MAGDOS_3,MAGDOS_4}. Also, as implied by the Pauli exclusion principle, the manipulation of the electronic occupation within the $\pi$ and $\pi^{*}$ bands alters the response to the electromagnetic (EM) perturbations. Gate-controlled optical absorption of graphene \cite{EXP_RES_2} and the broadband optical gain resulting from the inversion of the electronic population under femtosecond laser pulse irradiation \cite{INVP_1,INVP_2,INVP_3,INVP_4} are examples of altering the EM response of graphene via pushing its electronic occupation into steady and transient nonequilibrium states, respectively.
 
Modifying light absorption by electrical signals would integrate optics and electronics, a long-sought goal in plasmonics \cite{Ozb06}. Breaking the spatial-temporal symmetries would also open up the possibility of rectifying the plasmonic current to convert light signals into directed electrical signals.\cite{Olbrich09,Rozhansky15} The directional symmetry is most directly broken by applying an electric field within the two-dimensional (2D) layer. This will modify the spectrum, alter the response of the system and induce non-linear and thermal effects, while the presence of electrical contacts can lead to Dyakonov-Shur instabilities \cite{Dyakonov,Tomadin,Otsuji}. Also, population inversion induced by optical pumping can lead to a negative total dynamic conductivity in graphene at THz/far-infrared frequencies paving the way towards graphene-based laser devices.\cite{Ryzhii07,Satou13}

In this work, we will investigate the interplay between the electrical conductivity and the plasmonic excitations in graphene samples by assuming a moderate electric flux across the sample. To present the essence of our work, we focus on the analysis of the linear, intravalley response of drifting $\pi$ electron gas to longitudinal EM perturbations,
\begin{equation}\label{LONGITUDINAL_EM_PERTURBATION}
	\bm{E}(\bm{r},t)=\bm{E}_{0}\, e^{i(\bm{q} \cdot \bm{r} -\omega t)} \quad\quad ; \bm{E}_{0} \parallel \bm{q}
\end{equation}

The paper is organized as follows. Sec. \ref{sec:LRT} contains the basics of the linear response theory of Dirac systems and its generalization to nonequilibrium systems. In Sec. \ref{sec:ANAP}, we present the analytical approximation valid for small drift velocities at finite doping. In Sec. \ref{sec:DPS}, a general discussion for doped systems is given, and in Sec. \ref{sec:GAIN}, we present the numerical results for the case of zero doping. We close with a summary and conclusions. The paper is supplemented by four appendices which provide details on the analytical calculations.

\section{Linear response of a driven Dirac system}\label{sec:LRT}
Within the random phase approximation \cite{Mahan}(RPA), the response of the $\pi$ electrons at equilibrium to longitudinal EM perturbations is mainly described by the intravalley dynamical polarization function of graphene \cite{FNDOPE,HWANG},
\begin{equation}\label{Dynpo}
	\begin{split}
		\Pi(\bm{q},\omega) = \frac{g_s g_v}{(2\pi)^2} \!\! & \sum_{s,s'=\pm} \int d^{2}\bm{k} \, \{ f_{s,s'}(\bm{k},\bm{q})
		\\
		&\quad \;\;\frac{n_{F}[E^{s'}\!(\bm{k}\!+\!\bm{q})]-n_{F}[E^{s}(\bm{k})]}{E^{s'}\!(\bm{k}\!+\!\bm{q})-E^{s}(\bm{k})-\hbar \omega-i0^{+}}\}
	\end{split}
\end{equation}
where $g_s(g_v)\!=\!2$ denotes the spin (valley) degeneracy, the prefactor $f_{s,s'}\!\!\left(\bm{k},\bm{q}\right)$ represents the band overlap integral and $E^{s}(\bm{k})$ describes the energy dispersion of the valence ($s=-1$) and  conduction ($s=1$) bands. Employing the tight-binding model, if accompanied by the Dirac cone approximation, yields $E^{s}(\bm{k})=s(3at_{0}/2)k$ together with, 
\begin{equation}\label{Tight-Binding}
	f_{s,s'}\!\!\left(\bm{k},\bm{q}\right)=\frac{1}{2}[1+ss' \, \frac{k+q\cos\!{(\theta_{\bm{k}}\!-\!\theta_{\bm{q}})}}{\left|\bm{k}\!+\!\bm{q}\right|}]\quad\quad
\end{equation}
where $a\approx0.142 nm$, $t_{0}\approx2.7 eV$ and $\bm{k}$ are respectively the carbon-carbon bond length, the nearest-neighbor hopping amplitude and the crystal momentum measured with respect to the Dirac points. In addition, $\theta_{\bm{k}(\bm{q})}$ is the angle between $\bm{k}$ ($\bm{q}$) and $\hat{e}_{x}$. The equilibrium electronic occupation is described by the Fermi-Dirac statistics, i.e.,
\begin{equation}\label{FERDIR}
	n_{F}[E]=\left[1+\exp{\big(\frac{E-E_{F}}{k_{B}T}}\big)\right]^{-1}
\end{equation}
where $E_{F}$ is the Fermi energy measured with respect to the neutrality point. The highest occupied eigen-states in the reciprocal space are located at circles centered at the Dirac points. The disk enclosed by such a circle is referred to as the Fermi disk with the Fermi wavevector, $k_F=2\left|E_{F}\right|/(3at_{0})$, being its radius.

In the presence of drift, the $\pi$ electron gas reaches a new equilibrium through gaining momentum and kinetic energy from the drain-source voltage, $V_{ds}$, and losing part of it via electron scattering mechanisms \cite{GUO,XLI}. Instead of finding the eigen-states and energy eigen-values of the Hamiltonian that includes $\bm{E}_{ds}$ (the local electric field corresponding to $V_{ds}$) and the sources of scattering, we adopt a semi-classical approach in which the electronic occupation is altered by the drain-source voltage, while the crystal Hamiltonian and consequently its DOS, remain intact. As a result of this approach, the drift-induced modification to the dynamical polarization of graphene, which is defined as the difference between the dynamical polarization in the presence of drift $\Pi^{\bullet}(\bm{q},\omega)$ by its no-drift counterpart $\Pi^{\circ}(\bm{q},\omega)$, is given by \footnote{In this work, the symbols super-scripted with a filled and hollow circle respectively correspond to the cases where the drift is present and absent. Any symbol without such superscripts implicitly corresponds to the no-drift case.},
\begin{equation}\label{Dynpo_Modification}
	\begin{split}
		\Delta\Pi(\bm{q},\omega) \! \cong \! \frac{g_s g_v}{(2\pi)^2} \!\! & \sum_{s,s'=\pm}\int d^{2}\bm{k} \, \{ f_{s,s'}(\bm{k},\bm{q})
		\\
		&\;\;\; \frac{\Delta n_{F}[E^{s'}\!(\bm{k}\!+\!\bm{q})]-\Delta n_{F}[E^{s}(\bm{k})]}{E^{s'}\!(\bm{k}\!+\!\bm{q})-E^{s}(\bm{k})-\hbar\omega-i0^{+}}\}
	\end{split}
\end{equation}
where $\Delta  n_{F}[E]\equiv n_{F}^{\bullet}[E]-n_{F}^{\circ}[E]$ denotes the drift-induced modification to the electronic occupation.

\section{Analytic approximation}\label{sec:ANAP}
In principle, the occupation function of the drifting electron gas can be obtained via solving the Boltzmann transport equation (BTE) \cite{XLI}; however, in order to avoid the complexities of solving the BTE, we resort to the phenomenological shifted Fermi disk model which describes the nonequilibrium occupation function of a drifting electron gas without the need for the details of the underlying scattering mechanisms \cite{SFD_1,SFD_2}. For a given shift of the Fermi disk from the Dirac point, $\bm{k}_{shift}$, the locus of the highest occupied states of the drifting electron gas with respect to the Dirac point, i.e. $\bm{k}=\bm{0}$, is given by,
\begin{equation}\label{HOS_LOCUS}
	k_{F}^{\bullet}=k_{F}\left\{\sqrt{1-[\frac{k_{shift}}{k_{F}}\sin{\theta_{\bm{k}}}]^{2}}-[\frac{k_{shift}}{k_{F}}\cos{\theta_{\bm{k}}}]\right\}
\end{equation}
where $k_{shift}\! < \! k_{F}$ is implied and the shift is assumed to be leftward, i.e. $\bm{E}_{ds} \! \parallel \! \hat{\bm{e}}_{x}$. We thus limit ourselves to the case of pure electron or hole transport and relegate the special case of doping-levels close to half-filling to Sec. \ref{sec:GAIN}. Within the low-temperature ($k_{B}T \! \ll \! \left|E_{F}\right|$) and low-drift ($k_{shift} \! \ll \! k_F$) regime, which is a relevant regime for usual doping levels and current densities, the occupation function of the drifting electron gas $n_{F}^{\bullet}[E]$ can be approximated via feeding $E_{F}^{\bullet}=E_{F}(k_{F}^{\bullet}/k_{F})$ from Eq. (\ref{HOS_LOCUS}) into the Fermi-Dirac occupation function. This yields,
\begin{equation}\label{FD_perturbation}
	\Delta n_{F}(E,\bm{k}) \cong -E_{F} \, [\frac{k_{shift}}{k_{F}}] \, \delta(E-E_{F}) \, \cos{\theta_{\bm{k}}}
\end{equation}
with $\delta$ being the Dirac delta function. Feeding Eq. (\ref{Dynpo_Modification}) with this spike-like $\Delta n_{F}(E,\bm{k})$, if accompanied by the Dirac cone approximation, yields an analytic expression for $\Delta\Pi(\bm{q},\omega)$ (see Appendix \ref{App:A}). For brevity, we present this analytic expression in terms of the dimensionless variables $\tilde{q}\equiv q/k_{F}$, $\tilde{\omega}\equiv \hbar\omega/E_{F}$ and $\tilde{\omega}^{\prime}\equiv \tilde{\omega}+i0^{+}$,
\begin{equation}\label{GRPH_CLOSED-FORM}
	\Delta\Pi(\bm{q},\omega) \cong \! \frac{D(E_{F})}{4} \, \frac{\bm{q} \cdot \bm{v}_{dr}}{q \, v_{F}} \left[-\frac{8\tilde{\omega}}{\tilde{q}}\! + \!\! \sum_{\alpha=\pm} \! \alpha \, F_{\alpha}(\tilde{q},\tilde{\omega})\right]
\end{equation}
where $F_{\alpha}(\tilde{q},\tilde{\omega})$ is a complex function defined in terms of $Z_{\alpha} \equiv (2-\alpha \tilde{\omega}^{\prime})/\tilde{q}$ and $W_{\alpha} \equiv \alpha \tilde{\omega}^{\prime}/\tilde{q}$ as below,
\begin{equation}\label{GRPH_F_ALPHA}
	F_{\alpha}(\tilde{q},\tilde{\omega}) \equiv  \tilde{q} \left[1-Z_{\alpha}^{2}\right] \sqrt{\frac{\left[1+W_{\alpha} Z_{\alpha}\right]^{2}}{\left[1-W_{\alpha}^{2}\right]  \left[1-Z_{\alpha}^{2}\right]}}
\end{equation}
with $v_{F} \equiv 3at_{0}/2\hbar \approx 10^{6} \, m/s$ being the velocity of Dirac Fermions. The factor $D(E_{F})\!=\!g_{s} g_{v}\left|E_{F}\right|/[2\pi (\hbar v_{F})^{2}]$ is the DOS of the Dirac cones at $E \! = \! E_{F}$ and the symmetry-breaking role of the electric current is manifested as the inner product of the wavevector $\bm{q}$ and the drift velocity of the electron (hole) gas, $\bm{v}_{dr}=sgn[E_{F}] \, v_{F}\, (\bm{k}_{shift}/k_{F})$.
\begin{figure}
	\begin{flushleft}
		\includegraphics[clip,width=8.6 cm]{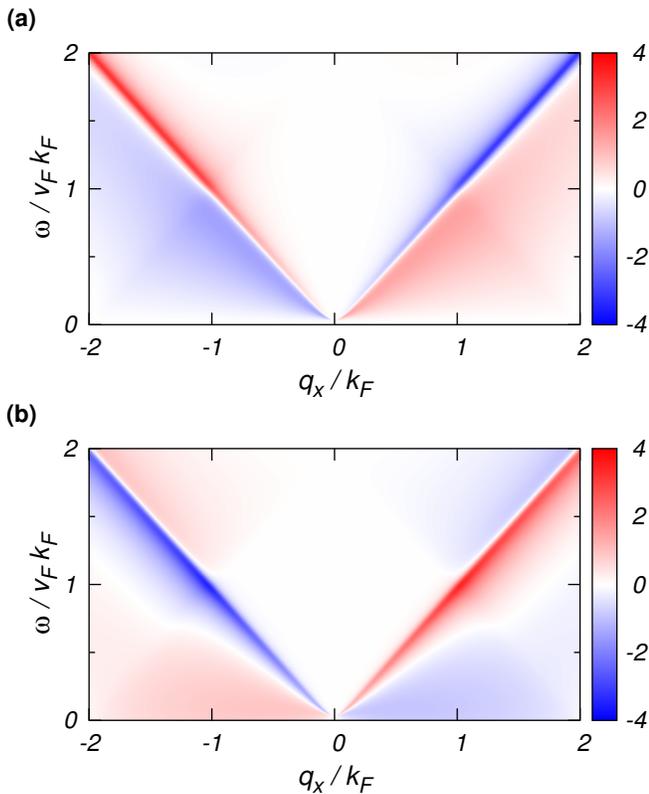} \\
		\caption{(Color online) Color-mapped values of {\bf (a)} the real and {\bf (b)} imaginary parts of $\Delta\Pi(\bm{q},\omega)$ obtained from Eq. (\ref{GRPH_CLOSED-FORM}) and presented in units of $(k_{shift}/k_{F}) D(E_{F})$. The positive and negative $q_{x}$ axes respectively correspond to the cases where $\bm{q}$ is anti-parallel ($\theta_{\bm{q}}=0^{\circ}$) and parallel ($\theta_{\bm{q}}=180^{\circ}$) to the drift velocity. The computed $\Delta\Pi(\bm{q},\omega)$ values are corrected according to the Mermin's approach \cite{Mermin} for a phenomenological scattering rate of $\hbar/\tau=5 \, meV$ (see Appendix \ref{App:B}).}
		\label{DP_MODIFICATION_PLOT}
	\end{flushleft}
\end{figure}
The analytic expression for $\Delta\Pi(\bm{q},\omega)$ given by Eq. (\ref{GRPH_CLOSED-FORM}) is the main result of this work and shown in Fig. \ref{DP_MODIFICATION_PLOT}. It conforms with the following principles:\\*i) {\it Real charge response.} It satisfies the below condition which guarantees a non-imaginary charge response,
\begin{equation}\label{THE_CONDITION}
	\Delta\Pi(-\bm{q},-\omega) =[\Delta\Pi(\bm{q},\omega)]^{*}
\end{equation}
ii) {\it Causality.} Since the integrand of $\Delta\Pi$ has no poles in the upper half-plane of the complex frequency space, the real and imaginary parts of the analytic expression for $\Delta\Pi(\bm{q},\omega)$ are automatically correlated through the Kramers-Kronig (KK) relations,
\begin{equation}\label{KK}
	\mathcal{P}\int\frac{\Delta\Pi(\bm{q},\omega')}{\omega'-\omega} \, d\omega'=i\pi\Delta\Pi(\bm{q},\omega)
\end{equation}
iii) {\it The f-sum rule.} The general f-sum rule for a bipartite tight-binding model \cite{STAUBER_10} implies,
\begin{equation}\label{F_SUM_RULE}
	\int\! \Im{[\Delta\Pi(\bm{q},\omega)]} \, \omega \, d\omega \, \propto \int \!\!\! \ E^{\pm}\!(\bm{k}) \, \Delta n_{F}[E^{\pm}\!(\bm{k})] \, d^{2}\bm{k}
\end{equation}
where the right-hand side (RHS) represents the drift-induced modification to the kinetic energy of the $\pi$ electron gas. The low-drift approximation for $\Delta n_{F}(E,\bm{k})$ given by Eq. (\ref{FD_perturbation}) does not alter the kinetic energy of the electron gas, i.e. the RHS vanishes. Within the same level of approximation, the analytic expression for $\Im{[\Delta\Pi]}$ given by Eq. (\ref{GRPH_CLOSED-FORM}) is an even function of $\omega$ and yields a vanishing LHS, thereby satisfying the sum rule. 
\begin{figure}
	\begin{center}
		\includegraphics[clip,width=8.4 cm]{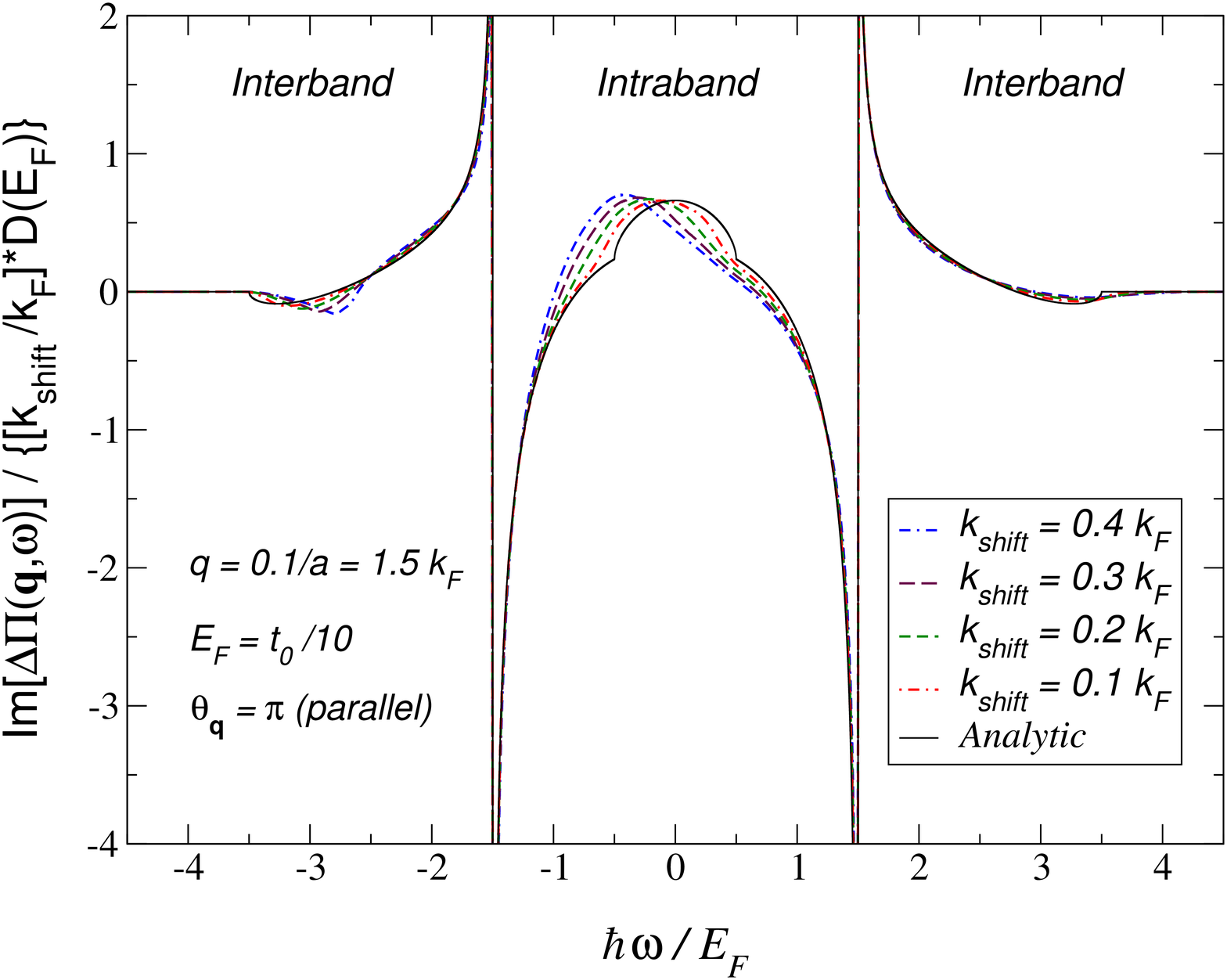} \\
		\caption{(Color online) The numeric-analytic comparison of $\Im{[\Delta\Pi(\bm{q},\omega)]}$ for Dirac Fermions indicates that the analytic expression given by Eq. (\ref{GRPH_CLOSED-FORM}) becomes more accurate for smaller amounts of $(k_{shift}/k_{F})$. The numerical evaluation of $\Delta\Pi$ is performed based on $\Delta n_{F}$ values computed from Eq. (\ref{HOS_LOCUS}), and the results are normalized by $(k_{shift}/k_{F}) \times D(E_{F})$.}
		\label{DP_COMPARE}
	\end{center}
\end{figure}

It is worthy to note that the analytic expression for $\Delta\Pi$ presented here contains only the terms that are linear in $k_{shift}/k_{F}$. The response for arbitrary drift velocity can be obtained numerically. In Fig. \ref{DP_COMPARE}, the numerical result for the imaginary part of $\Delta\Pi$ is shown for several (large) drift velocities at a fixed wave number $q=1.5k_{F}$ in the direction of the drift.

 Let us finally note that within the framework of the shifted Fermi disk model, an exact analytic expression for the drift-induced intrasubband dynamical polarization of two-dimensional electron gas (2DEG) \cite{Stern} is obtainable whose validity extends beyond the low-temperature and low-drift regime (see Appendix \ref{App:C}),
\begin{equation}\label{2DEG_DYNPO_MOD_EXACT}
	\Pi^{\bullet}(\bm{q},\omega)=\Pi ^{\circ}(\bm{q},\omega-(\hbar/m_{e}^{*})[\bm{k}_{shift}\cdot\bm{q}]) \;
\end{equation}
which suggests that the drifting 2DEG responds to the EM perturbation with a Doppler-shifted frequency.

\section{Discussion for doped systems}\label{sec:DPS}
\subsection{The static limit}\label{subsec:DPS-STATIC}
Within the low-drift and low-temperature regime, the analytic expression for the drift-induced modification to the intravalley static polarization of the $\pi$ electron gas in graphene can be obtained from Eq. (\ref{GRPH_CLOSED-FORM}) via setting $\omega=0$,
\begin{equation}\label{GRPH_STATIC}
	\Delta\Pi(\bm{q},\omega=0) \cong i \, D(E_{F}) \, \frac{\hat{\bm{q}} \cdot \bm{v}_{dr}}{v_{F}} \, \Re{[\sqrt{1 \! - \! \left[\!\frac{q}{2k_{F}}\!\right]^{2}}]}\;.
\end{equation}
One of the notable consequences of such a modification is the emergence of a drift-induced asymmetry in the Friedel oscillations (FO) \cite{FNDOPE}. At far enough distances from the charged impurity atom, i.e. $r \, \gsim \, k_{F}^{-1}$, the modification to the FO in the presence of drift is described by,
\begin{equation}\label{FO_MOD}
	\frac{\Delta n_{ind}(\bm{r})}{n_{s}} \cong \alpha_{f} \, \frac{c}{v_{F}} \, \frac{sgn[Q]}{\pi \kappa_{0}^{2}} \, \frac{\bm{v}_{dr} \cdot \hat{\bm{r}}}{v_{F}} \, \frac{\sin{(2k_{F}r)}}{(k_{F}r)^{2}}
\end{equation}
where $n_{s}\!=\! k_{F}^{2}/\pi$ is the density of dopant electrons or holes, $Q$ is the charge of the impurity atom, $\bm{r}$ is the in-plane position vector, $c$ is the phase velocity of light in vacuum, $\alpha_{f}\! =\! e^{2}/4 \pi \varepsilon_{0}\hbar c$ (mks units) denotes the fine structure constant and $\kappa_{0}$ is the background dielectric constant \cite{FNDOPE}. Even though the drift-induced modification to the intervalley static polarization can be comparable to its intravalley counterpart, its contribution to the FO is negligible at far enough distances from the impurity atom, i.e. $r \, \gsim \, k_{F}^{-1}$. This is because the relatively large valley separation leads to rapidly-oscillating terms in the summation yielding the intervalley contribution.
\subsection{The local plasmonic limit}\label{subsec:DPS-PLASMONIC}
\begin{figure}
	\begin{center}
		\includegraphics[clip,width=8.4cm]{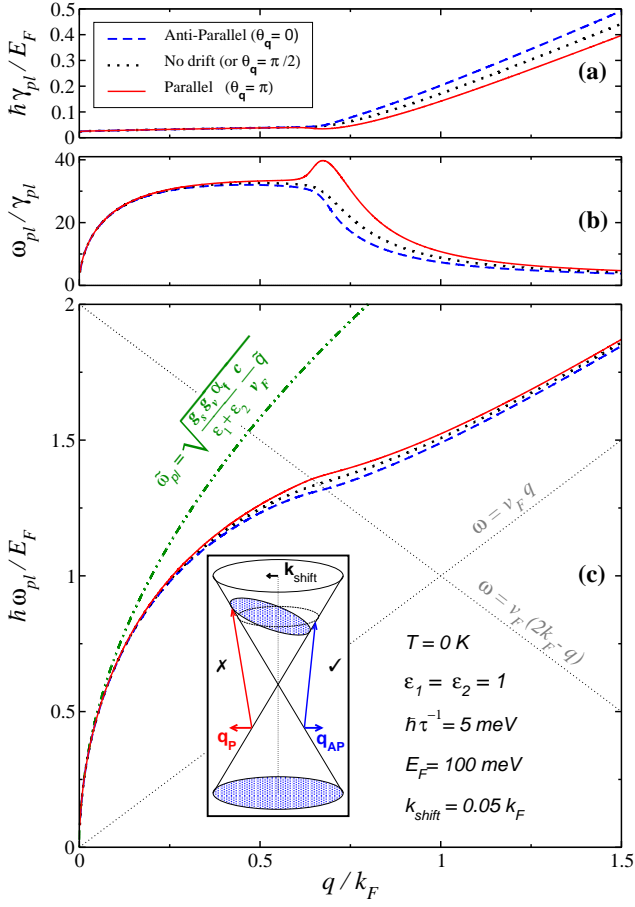}\\
		\caption{(Color online) The computed {\bf (a)} decay rate, {\bf (b)} $\omega_{pl}/\gamma_{pl}$ ratio and {\bf (c)} frequency of the TM-SPP modes of a suspended graphene channel, propagating parallel (red solid curves) and anti-parallel (blue dashed) to the drift velocity, are compared with their no-drift counterpart (black dotted curves). The agreement with the $\sqrt{q}$ behavior predicted by the local approximation (green dash-dotted) in the $q \ll k_{F}$ limit can be seen. The impact of the disorder-induced electron scattering is taken into account by the Mermin's approach using a phenomenological scattering rate of $\hbar/\tau=5 \, meV$.}
		\label{TM-SPP}
	\end{center}
\end{figure}
Within the local limit, i.e. $\tilde{q} \ll \left|\tilde{\omega}\right| \ll 1$, the dynamical polarization of doped $\pi$ electron gas is given by \cite{FNDOPE},
\begin{equation}\label{DP_LOCAL}
	\Pi(\bm{q},\omega) \cong \frac{1}{2} \, D(E_{F}) \left(\frac{\tilde{q}}{\tilde{\omega}}\right)^{2}
\end{equation}
and the expression for $\Delta\Pi$ given by Eq. (\ref{GRPH_CLOSED-FORM}) reduces to,
\begin{equation}\label{DP_MODIFICATION_LOCAL}
	\Delta\Pi(\bm{q},\omega) \cong \frac{1}{4} \, \frac{\bm{q} \cdot \bm{v}_{dr}}{q \, v_{F}} \, D(E_{F}) \left(\frac{\tilde{q}}{\tilde{\omega}}\right)^{3}
\end{equation}
This translates into the drift-induced modification to the Drude weight \cite{Stauber_TWISTED_BLG} given as follows,
\begin{equation}\label{DRUDE_WEIGHT_MODIFICATION_LOCAL}
	\Delta D=\frac{{\bm{q}} \, \cdot \bm{v}_{dr}}{2\omega}D_{0}
\end{equation}
where $D_{0} \! = \! e^{2} \left|E_F\right|/(\pi \hbar^{2})$ is the bare Drude weight of the dopant electrons or holes. The local-limit expression for the dynamical polarization given by Eq. (\ref{DP_LOCAL}) yields a $\sqrt{q}$ dependence for the TM-SPP frequency which is shown in Fig. \ref{TM-SPP}(c). On the other hand, an additional acoustic branch with a linear dispersion, i.e. $\omega\sim v_{s}q$, emerges in a double-layer \cite{Sarma_DLG,Stauber_DLG,Rosario_DLG} or gated \cite{Svintsov_Gated_Pl} graphene system. The drift-induced change to the Drude weight can usually be neglected for the optical branch, i.e., $\omega\sim\sqrt{q}$. For the acoustic branch, however, this correction might become observable if the sound-velocity, $v_{s}$, is comparable to the drift velocity $v_{dr}$.

\subsection{Modified plasmon dispersion and damping}\label{subsec:MPD}
A two-dimensional (2D) electron gas confined between two dielectric media of dielectric constants $\epsilon_{1}$ and $\epsilon_{2}$ supports transverse magnetic surface plasmon polariton (TM-SPP) modes whose dispersion is yielded by the solutions of the following equation \cite{Stern,Marinko,Bludov},
\begin{equation}\label{TM-SPP_DISPERSION}
	\frac{\epsilon_{1}}{\sqrt{1-\epsilon_{1}(\frac{\omega}{qc})^{2}}}+\frac{\epsilon_{2}}{\sqrt{1-\epsilon_{2}(\frac{\omega}{qc})^{2}}}= \frac{2\alpha_{f}hc}{q} \; \Pi(\bm{q},\omega)
\end{equation}

The retardation region is defined as the region in the $(q,\omega)$ plane near the dispersion of light \cite{Marinko,Tobias_toprev}. Since the EM fields corresponding to the modes located in the retardation region are poorly-localized to the graphene sheet \cite{Marinko,Bludov}, we discuss the effects of drift out of this regime, i.e. $\tilde{q}/\tilde{\omega} \gg v_{F}/c$, where the LHS of Eq. (\ref{TM-SPP_DISPERSION}) reduces to $\epsilon_{1}+\epsilon_{2}$.  To simplify our study and block the plasmon damping pathways such as the plasmon decay into the intrinsic ($\hbar\omega \approx 195 \, meV$) and extrinsic optical phonon modes within the frequency range of our interest ($\hbar\omega \, \lsim \, 2E_{F}$) \cite{Yan,Buljan}, we assume the Fermi energy to be low enough ($E_{F} \lsim \, 0.1 eV$) and we limit our study to the case of a nonpolar substrate. Comparing the dispersion relation in the absence and presence of drift at a fixed $q$ yields,
\begin{equation}\label{DP_STAY_FIXED}
	\Pi^{\bullet}(\bm{q},\omega_{pl}^{\bullet}(\bm{q}))= \Pi^{\circ}(\bm{q},\omega_{pl}^{\circ}(\bm{q}))
\end{equation}
which implies that the electric flux modifies the TM-SPP frequency. Within the low-drift regime, Eq. (\ref{DP_STAY_FIXED}) yields the drift-induced modification to the TM-SPP frequency $\omega_{pl}(\bm{q})$ and decay rate $\gamma_{pl}(\bm{q})$,
\begin{equation}\label{TM-SPP_MODIFICATION}
	\Delta \omega_{pl}(\bm{q})-i\Delta \gamma_{pl}(\bm{q}) \cong -\frac{\Delta\Pi(\bm{q},\omega_{pl}^{\circ}(\bm{q}))}{\left[\frac{\partial \Pi(\bm{q},\omega)}{\partial \omega}\right]_{\omega=\omega_{pl}^{\circ}(\bm{q})}}
\end{equation}
Within the local limit, Eq. (\ref{TM-SPP_MODIFICATION}) yields an expression for the drift-induced modification to the frequency of the TM-SPP modes that is valid within the local limit,
\begin{equation}\label{LOCAL_TM-SPP_FREQUENCY_MODIFICATION}
	\Delta \omega_{pl}(\bm{q}) \cong \frac{1}{4} \, \bm{q} \cdot \bm{v}_{dr} \quad\quad\;\; ;q \ll k_{F}
\end{equation}
This Doppler-like frequency modification suggests that in the local limit the plasmonic charge density excitations are partially dragged by the drifting $\pi$ electron (hole) gas. As it is shown in Fig. \ref{TM-SPP}c, the presence of drift causes the TM-SPP dispersion of $\pi$ electron gas to split into two branches each of which corresponds to the TM-SPP modes propagating parallel ({\bf P}), i.e. $\theta_{\bm{q}}=\pi$, and anti-parallel ({\bf AP}), i.e. $\theta_{\bm{q}}=0$, to the drift velocity. This splitting can be inferred from the following expression,
\begin{equation}\label{PROPORTIONALITY_TM-SPP_FREQUENCY_MODIFICATION}
	\Delta \omega_{pl}(\bm{q})-i\Delta \gamma_{pl}(\bm{q}) = [\bm{q} \cdot \bm{v}_{dr}] \, \Upsilon(q)
\end{equation}
where $\Re{[\Upsilon(q)]},\Im{[\Upsilon(q)]}>0$. The computed TM-SPP branches that are presented in Fig. \ref{TM-SPP}(c) suggest that the drift-induced frequency splitting, i.e. $2 q v_{dr} \Re{[\Upsilon(q)]}$, reaches its maximum near $q=q_{c}$ with $q_{c}$ denoting the onset of the Landau damping \cite{Liu,FNDOPE} (see Appendix \ref{App:D}).

Moreover, according to Fig. \ref{TM-SPP}(a), the TM-SPP mode propagating parallel (anti-parallel) to the drift velocity has a longer (shorter) lifetime comparing to the case in which the drift is absent. As it is implied by Fig. \ref{TM-SPP}(b), such a drift-induced change in the propagation length, if measured in units of the mode wavelength, reaches its maximum for the modes with $q \approx q_{c}$. As a result, the short (few-wavelength) propagation length of plasmons, which is the main challenge of graphene plasmonics \cite{Tassin}, can be increased via the application of a drain-source voltage. This conclusion, nevertheless, is based on the assumption that the device temperature is not affected by the presence of the drift. Otherwise, to assess the overall drift-induced change in the TM-SPP propagation length, the increase in the temperature-induced plasmon damping \cite{Yan,Tassin,Sarma_Li} resulting from the Joule heating of the current-carrying device should be taken into account. The higher decay rate for the {\bf AP} branch, suggests the possibility of the application of DC current as a plasmonic brake to establish a one-way EM waveguide \cite{Engheta,Genov,Erping}.

As was noted in Sec. \ref{sec:ANAP}, only the terms linear in $k_{shift}/k_{F}$ are retained in the analytic expression. Such an approximation produces unphysical results within a tiny neighborhood of  the onset of Landau damping, i.e. $q_{c}\pm\delta q$ and $\omega_{pl}(q_{c})\pm\delta\omega$ where $\delta \omega=v_{F}\delta q \propto \bm{v}_{dr} \cdot \bm{q}$. The slight dip in the decay rate curve and the exaggerated peak in the $\omega_{pl}/\gamma_{pl}$ curve of the {\bf P} plasmons, which are respectively presented by Fig. \ref{TM-SPP}(a) and Fig. \ref{TM-SPP}(b), are the inevitable consequences of such an approximation. To remedy this shortcoming, the terms proportional to $(k_{shift}/k_{F})^{n\geq2}$ should be derived and included in the analytic expression.

\section{Zero doping and plasmon gain}\label{sec:GAIN}
\begin{figure}
	\begin{center}
		\includegraphics[clip,width=8cm]{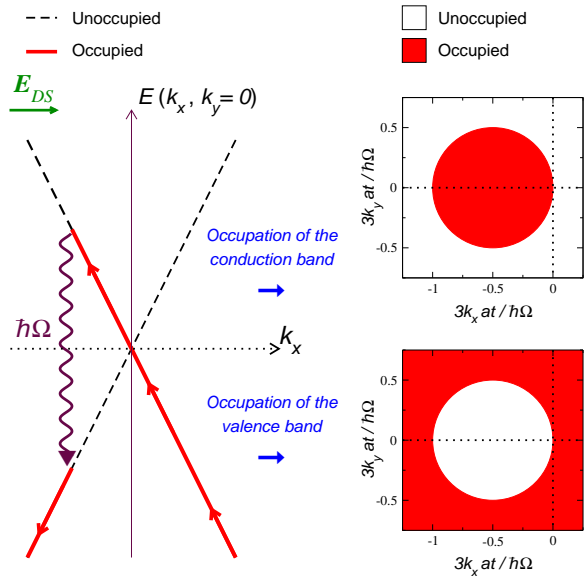}\\
		\caption{(Color online) Depiction of the $\bm{k}$-space dynamics \cite{A&M} of the $\pi$ electron gas (the straight arrows along the cross-section of the Dirac cones with the $k_{y}=0$ plane) in an undoped graphene channel subjected to a drain-source voltage. The drifting occupants of the valence band whose group velocity has an opposite component to $\bm{E}_{ds}$ migrate to the conduction band through the Dirac crossing. Ultimately, the migrants backscatter to the valence band after losing a quantum of energy $\hbar \Omega$ via the emission of a phonon mode (wavy arrow).}
		\label{k_DYNAMICS}
	\end{center}
\end{figure}

Aside from the mode which was predicted when including the vertex corrections \cite{Gangadharaiah08}, undoped graphene does not support any TM-SPP modes at $T=0K$ within the RPA. \cite{Sarma_Li} Here, we numerically show that a high enough drain-source voltage along an undoped graphene channel enables the channel to support specific TM-SPP modes, even for the purely hypothetical case of $T=0 K$. More importantly, the numerical results indicate the possibility of the emission of the low-energy ($\hbar\omega \, \lsim \, 30 \, meV$) and long wavelength plasmons. Similar proposals can be found in Refs. \onlinecite{Ryzhii07,Satou13} and references therein.

The crossing nature of the conduction and valence bands in graphene obligates the drifting electrons in the valence band to move up to the conduction band because $\nabla_{\bm{k}}E^{s}(\bm{k})$, which is semi-classically interpreted as the group velocity, is not well-defined at $\bm{k}=\bm{0}$ for a single Dirac cone. That is, a drifting electron passing through the neutrality point must travel to the other band. As is shown by Fig. \ref{k_DYNAMICS}, if the drain-source voltage is high enough, the migrant electrons lose a quantum of energy $\hbar \Omega$ by emitting a phonon mode and backscatter to the valence band.

The intrinsic high-field transport properties of metallic single-wall carbon nanotubes (SWCNTs) \cite{Dekker} is one piece of evidence that proves that such a transport model is a physically relevant one. In each of the two bands, i.e. $s=\pm1$, the nonequilibrium electronic occupation can be modeled by a $\theta_{\bm{k}}$-dependent Fermi energy as illustrated in Fig. \ref{k_DYNAMICS} and described by the following expression,
\begin{equation}\label{UNDOPED_DRIFTING_MU}
	E_{F,s}^{\,\bullet}(\theta_{\bm{k}})=\frac{s}{4}\,\hbar\Omega \left[\left|\cos{\theta_{\bm{k}}}\right|-\cos{\theta_{\bm{k}}}\right]\quad\quad ;s=\pm1\;.
\end{equation}
Regarding the 1D nature of the conical subbands of the metallic SWCNT,\cite{Dekker} the model given by Eq. (\ref{UNDOPED_DRIFTING_MU}) may not be the best one to describe the similar situation within the 2D Dirac cones of graphene. However, any other possible model should also allow for nearly vertical, i.e. $qa \ll 1$, electron-hole recombination processes, which is a guarantee for the generation of the long wavelength interband plasmons.

\begin{figure}
	\begin{flushleft}
		\includegraphics[clip,width=8.6 cm]{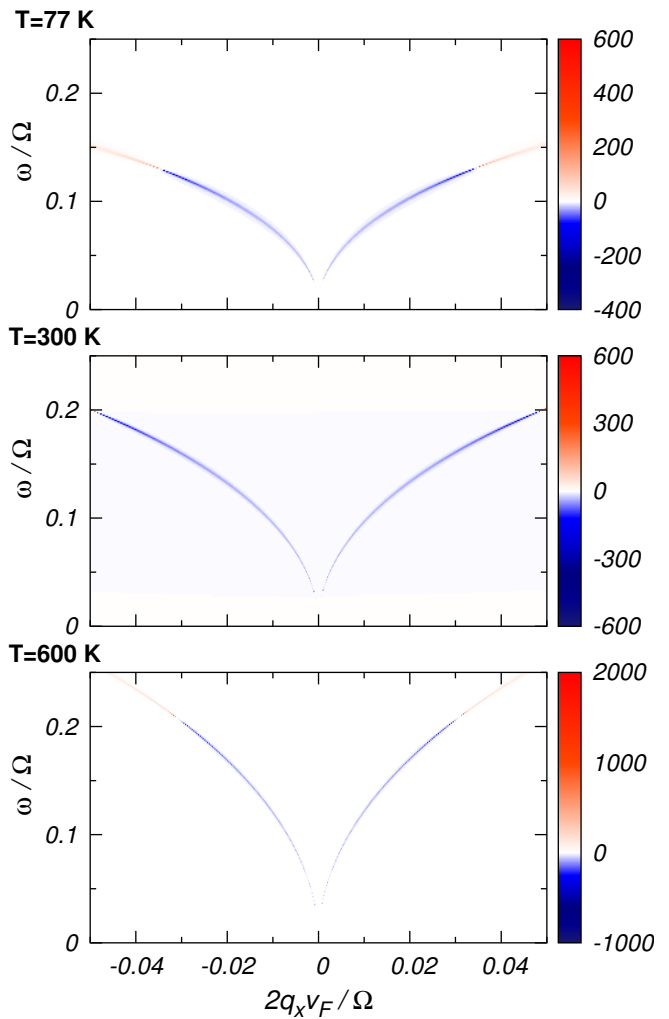} \\
		\caption{(Color online) Color-mapped values of the energy loss function, $S(\bm{q},\omega)$, of the $\pi$ electrons in a suspended and undoped graphene channel along which a high drain-source voltage is applied. The negative energy loss for the TM-SPP modes with $q_{min} \leq q \leq q_{max}$ indicates the possibility of the amplification of these modes through the current saturation mechanism ($\hbar\Omega \approx 149 \, meV$). The negative and positive $q_{x}$ axes respectively correspond to the {\bf P} and {\bf AP} cases.}
		\label{Plasmon_emission}
	\end{flushleft}
\end{figure}

Here, the dynamical polarization of the $\pi$ electron gas in a current-saturated and intrinsic graphene channel is approximated by feeding the response function with the nonequilibrium Fermi energy given by Eq. (\ref{UNDOPED_DRIFTING_MU}). In order to identify the well-localized TM-SPP modes in the $(\bm{q}.\omega)$ plane, we rely on the energy loss function $S(\bm{q},\omega)$, which is a measure of the spectral intensity of these modes \cite{Tobias_toprev},
\begin{equation}\label{LOSS_FUNCTION}
	S(\bm{q},\omega)=-\Im{\left[\left(1-\frac{2\alpha_{f}hc}{[\epsilon_{1}+\epsilon_{2}] \, q} \; \Pi(\bm{q},\omega)\right)^{-1}\right]}
\end{equation}
The computed energy loss, which is presented in Fig. \ref{Plasmon_emission}, suggests that: i) the presence of electric current causes certain TM-SPP modes to emerge (which we refer to as the ``drift-born'' modes). ii) The electric current does not introduce any asymmetry to the response. The latter is due to the peculiar nonequilibrium occupation of the drifting $\pi$ electrons. The most important feature of the nonequilibrium response presented in Fig. \ref{Plasmon_emission} is the negative energy loss of TM-SPP modes with $q_{min} \leq q \leq q_{max}$. Accordingly, we propose the use of the current-saturation in a nearly-undoped graphene channel as a mechanism for the amplification of THz plasmons.

Since the electron-electron ($e$-$e$) interactions in the $\pi$ electron gas in graphene become significant for very low densities of dopant electrons, \cite{Shankar,Kotov} it is necessary to discuss whether the validity of the results presented for the undoped case is challenged by the $e$-$e$ interactions. Experimentally \cite{Mayorov12} as well as theoretically \cite{Ulybyshev13}, it has been established that there is no gap opening due to chiral symmetry breaking even at the largest effective coupling constant present in suspended graphene ($\alpha_{g}=2.2$). However, strong Fermi velocity renormalisation takes place for low densities of the order $n_{s}\sim10^{10}$cm$^{-2}$ up to a factor of three.\cite{Siegel,Elias} To a first approximation, this effect can be taken into account by using the renormalized Fermi velocity instead of the bare one (i.e., $v_F$). Therefore, the $e$-$e$ interactions would not hinder the plasmon amplification mechanism proposed here but rather would modify the quantitative aspects of the numerical results we presented in the non-interacting picture (Fig. \ref{Plasmon_emission}), and additional work is needed to clarify this issue.

The electron and hole puddles, i.e. the spatial fluctuations in the Fermi energy, set a technical barrier to achieving a uniform neutrality along the graphene channel.\cite{Yacoby} Fortunately, it has been shown experimentally that these charge puddles can be substantially reduced on a hexagonal boron nitride (hBN) substrate,\cite{LeRoy,Crommie} thereby leaving some possibility for the plasmon amplification mechanism proposed in this work. However, theoretical investigations \cite{Giovannetti,Slawinska,Adam} and experimental measurements \cite{Hunt,Woods} suggest that a proper crystallographic alignment of graphene with the hBN leads to the local breaking of the sublattice symmetry, thus opening a sizable band gap at the Dirac point. There are two grounds on which it can be shown that the use of hBN substrate does not necessarily induce any band gap: i) The sublattice symmetry in graphene can only be broken for specific relative rotation angles between the crystals, and this is why several works failed to detect such a gap. \cite{Dean,Shepard,LeRoy,Crommie} ii) Even if graphene is properly aligned with the hBN crystal so that the band gap emerges, placing an additional hBN crystal on top of graphene would kill the commensurate state and recover the sublattice symmetry.\cite{Mayorov,Woods} Therefore, such an unfavorable gap, which obstructs the proposed plasmon amplification mechanism, can be feasibly avoided by a proper encapsulation of graphene with hBN.

Regarding the high temperature of a graphene channel within the current-saturation regime,\cite{HEINZ} it is necessary to incorporate the effects of temperature into the evaluation of the energy loss function, and as is shown in Fig. \ref{Plasmon_emission}, the negative energy loss persists at high temperatures. 

\section{Summary and conclusions}\label{sec:S&C}
We have discussed the dynamical response of a Dirac system subjected to a source-drain current. This was done by considering the nonequilibrium distribution function and feeding it into the well-known Lindhard function \cite{Dressel}. By this, we were able to obtain closed-form expressions within the low-drift limit and analyzed the nonequilibrium response. Since the f-sum rule is obeyed in this limit, our approximation can be regarded as quasi-equilibrium. However, the sum rule does not hold anymore for the case of $k_{shift} \sim k_{F}$, i.e., for systems out of equilibrium.

For doped systems, the asymmetric response of the drifting $\pi$ electron gas was discussed in the static and dynamical limit, especially commenting on the modified plasmon dispersion and damping rates. For a neutral system, where the external electric field does not induce an asymmetric response due to particle-hole symmetry, numerical results of the energy loss function were presented and a plasmon gain region found which persists even at high temperatures such as $T=600$K. This result may be relevant and lead to potential applications based on ultra-clean encapsulated graphene samples.
 
We finally discuss the limitations of the shifted Fermi sea model. Relaxation processes in graphene are known to be fast\textemdash{}in particular, $e$-$e$ relaxation tends to equilibrate the system within femtoseconds. Our model is therefore mainly valid in the analytical limit $k_{shift}/k_{F}\ll1$ which is reaffirmed by the fact that the sum-rule holds in this case of quasi-equilibrium. However, note that the Doppler-like transformation implied by Eq. (\ref{2DEG_DYNPO_MOD_EXACT}) is in concordance with the experimental measurements on the drifting 2DEG system \cite{HUGHES}. This can be regarded as supporting evidence for our approximate treatment of the EM response of a driven electron gas.

Several extensions are possible such as discussing the response of a gapped system, multi-layer systems and comparing our results with non-linear response functions.\cite{Mikhailov14}

\begin{acknowledgments}
We thank Daniel R. Mason and Guillermo G\'omez Santos for useful discussions. This work has been supported by the National Research Foundation of Korea (National Honor Scientist Program, 2010-0020414) and by the Ministerio de Econom{\'i}a y Competitividad (FIS2013-48048-P, FIS2014-57432-P).
\end{acknowledgments}

\appendix
\section{Derivation of the drift-induced modification to the dynamical polarization}\label{App:A}
\begin{figure}
	\begin{center}
		\includegraphics[clip,width=8cm]{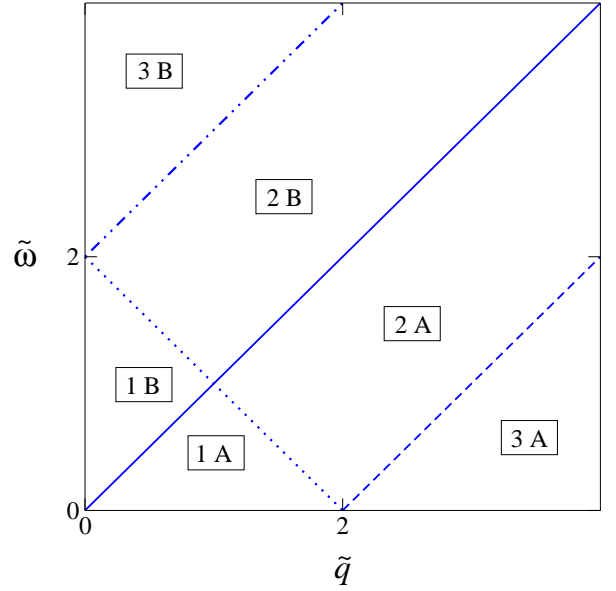}\\
		\caption{(Color online) The analytic solution for $\Delta\Pi(\bm{q},\omega)$ can be specified in each of the six regions in the $(\tilde{q},\tilde{\omega})$ plane which are outlined by the straight lines of $\tilde{\omega}=\tilde{q}$ (solid), $\tilde{\omega}=2+\tilde{q}$ (dash-dotted), $\tilde{\omega}=2-\tilde{q}$ (dotted) and $\tilde{\omega}=\tilde{q}-2$ (dashed).}
		\label{REGIONS}
	\end{center}
\end{figure}

For an n(a p)-doped graphene sample, the contribution of the occupied (empty) eigen-states of the conduction (valence) band to the dynamical polarization of the $\pi$ electron (hole) gas can be separated out as follows\cite{FNDOPE}:
\begin{equation}\label{DYNPO_RESOLVE}
	\Pi(\bm{q},\omega)-\Pi^{E_{F}=0}(\bm{q},\omega)=D(E_{F})\sum_{\alpha=\pm}U_{\alpha}(\bm{q},\omega)
\end{equation}
At $T=0$, the complex function $U_{\alpha}(\bm{q},\omega)$ is given by
\begin{equation}\label{DYNPO_DOPE_COMPONENT}
	U_{\alpha}(\bm{q},\omega)=\int_{0}^{2\pi} \frac{d\theta_{\bm{k}}}{2\pi} \; \int_{0}^{k_{F}} \frac{dk}{k_{F}} \; \Lambda_{\alpha}(\bm{k},\bm{q},\omega)
\end{equation}\;,
where $\Lambda_{\alpha}(\bm{k},\bm{q},\omega)$ is defined as follows:
\begin{equation}\label{DYNPO_DOPE_INTEGRAND}
	\Lambda_{\alpha}\equiv v_{F} k\sum_{\beta=\pm}\frac{f_{+,\alpha}\!\!\left(\bm{k},\bm{q}\right)}{\beta\omega \! + \! v_{F}\left[k \! - \! \alpha\left|\bm{k}\!+\!\bm{q}\right|\right] \! + \! i\beta 0^{+}}
\end{equation}
However, in a case where the rotational symmetry around the Dirac point is broken (e.g., a $\theta_{\bm{k}}$-dependent Fermi wavevector), the formalism given by Eqs. (\ref{DYNPO_RESOLVE}), (\ref{DYNPO_DOPE_COMPONENT}) and (\ref{DYNPO_DOPE_INTEGRAND}) does not satisfy the condition in Eq. (10). It is straightforward to derive the general formalism that holds for the non-symmetric case directly from Eq. (\ref{Dynpo}). The appearance of $\beta \bm{q}$ instead of $\bm{q}$ is the feature which distinguishes the general formalism from the old one,
\begin{equation}\label{DYNPO_DOPE_INTEGRAND_INVARIANT}
	\Lambda_{\alpha}^{\prime} \equiv  v_{F} k \sum_{\beta=\pm}\!\frac{f_{+,\alpha}\!\!\left(\bm{k},\beta\bm{q}\right)}{\beta\omega \! + \! v_{F}\left[k \! - \! \alpha\left|\bm{k}\!+\!\beta\bm{q}\right|\right] \! + \! i\beta 0^{+}}\;.
\end{equation}
Hence, the drift-induced modification to the dynamical polarization is given by,
\begin{equation}\label{DYNPO_MOD_APPENDIX}
	\Delta\Pi(\bm{q},\omega) = \frac{D(E_{F})}{2\pi k_{F}} \int_{0}^{2\pi} \!\! d\theta_{\bm{k}} \int_{k_{F}}^{k_{F}^{\bullet}} \!\! \left\{\Lambda_{+}^{\prime}+\Lambda_{-}^{\prime}\right\} dk\;.
\end{equation}
Within the low-drift regime, Eq. (\ref{HOS_LOCUS}) reduces to
\begin{equation}\label{kF_perturbation}
	k_{F}^{\bullet}\cong k_{F}\left\{1-\frac{ k_{shift}}{k_{F}}\cos{\theta_{\bm{k}}}\right\} \quad\quad\; ;\frac{ k_{shift}}{k_{F}}\ll 1\;.
\end{equation}
Regarding the small drift-induced perturbation to the Fermi wavevector, i.e. $\left|k_{F}^{\bullet}-k_{F}\right| \ll k_{F}$, the $k$-integral can be approximated as follows: 
\begin{equation}\label{DYNPO_APPROXIMATION}
	\int_{k_{F}}^{k_{F}-k_{shift}\cos{\theta_{\bm{k}}}} \!\!\!\!\! \Lambda_{\alpha}^{\prime}\; dk \cong  - \left[\Lambda_{\alpha}^{\prime}\right]_{k=k_{F}} k_{shift} \cos{\theta_{\bm{k}}}
\end{equation}
Substituting this result into Eq. (\ref{DYNPO_MOD_APPENDIX}) yields
\begin{equation}\label{DYNPO_DOPE_COMPONENT_MOD}
	\Delta\Pi(\bm{q},\omega) \cong \frac{D(E_{F})}{2\pi}\frac{k_{shift}}{k_{F}} \int_{0}^{2\pi} \! \left\{B_{+}+B_{-}\right\}d\theta_{\bm{k}}\;,
\end{equation}
where the integrand in Eq. (\ref{DYNPO_DOPE_COMPONENT_MOD}) is given by
\begin{equation}\label{DELTA_SELECTION}
	B_{\alpha}(\bm{q},\omega,\varphi)=- \left[\Lambda_{\alpha}^{\prime}(\bm{k},\bm{q},\omega)\right]_{k=k_{F}}\cos{(\varphi+\theta_{\bm{q}})}\;.
\end{equation}
The real part of the integral in Eq. (\ref{DYNPO_DOPE_COMPONENT_MOD}) can be obtained via applying the following integral identity:
\begin{equation}\label{INTEGRAL_IDENTITY_REAL}
	\int_{0}^{2\pi} \!\!\! \frac{d\varphi}{1-p\cos{\varphi}}=\, 2\pi \, \frac{\Theta[1-p^{2}]}{\sqrt{1-p^{2}}}
\end{equation}
with $p$ being a real number and $\Theta$ denotes the Heaviside's step function. We then arrive at the following expression,
\begin{equation}\label{DYNPO_SOLUTION_REAL}
	\Re{[\Delta\Pi(\bm{q},\omega)]}\cong A\,\left[\frac{8\tilde{\omega}}{\tilde{q}}+\sum_{\alpha=\pm}D_{\alpha}^{R}G_{\alpha}^{R}(\tilde{q},\tilde{\omega})\right]
\end{equation}
where the function  $G_{\alpha}^{R}(\tilde{q},\tilde{\omega})$ reads
\begin{equation}\label{DYNPO_FUNCTION_REAL}
	G_{\alpha}^{R}(\tilde{q},\tilde{\omega}) \!=\! \frac{\left|\tilde{\omega}(\tilde{\omega}-2\alpha)-\tilde{q}^{2}\right|[(\tilde{\omega}-2\alpha)^{2}-\tilde{q}^{2}]}{\tilde{q} \sqrt{(\tilde{\omega}^{2}-\tilde{q}^{2})[(\tilde{\omega}-2\alpha)^{2}-\tilde{q}^{2}]}}
\end{equation}
and the coefficient $D_{\alpha}^{R}$ is given by
\begin{equation}\label{DYNPO_COEFFICIENT_REAL}
	D_{\alpha}^{R}=+\alpha \, \Theta[(\tilde{\omega}^{2}-\tilde{q}^{2})\left\{(\tilde{\omega}-2\alpha)^{2}-\tilde{q}^{2}\right\}]\;.
\end{equation}
The following integral identity leads us to the imaginary part of the integral in Eq. (\ref{DYNPO_DOPE_COMPONENT_MOD}):
\begin{equation}\label{INTEGRAL_IDENTITY_IMAGINARY}
	\Im{\left[\int_{0}^{2\pi} \!\!\! \frac{N(\varphi) \, d\varphi}{M(\varphi)+i0^{\pm}}\right]}=\mp \pi \sum_{j=1}^{2}\left[\frac{N(\varphi)}{\left|\frac{dM(\varphi)}{d\varphi}\right|}\right]_{\varphi=\varphi_{j}}
\end{equation}
with $N(\varphi)$ and $M(\varphi)$ being two analytic functions within the range of $[0,2\pi]$, and $\varphi_{1,2}$ are the duet roots of $M(\varphi)$. The resulting expression for $\Im{[\Delta \Pi]}$ is as follows:
\begin{equation}\label{DYNPO_SOLUTION_IMAGINARY}
	\Im{[\Delta\Pi(\bm{q},\omega)]}\cong\frac{A}{\tilde{q}}\!\sum_{\alpha=\pm} \! D_{\alpha}^{I} \, G_{\alpha}^{I}(\tilde{q},\tilde{\omega}) \, sgn[\tilde{\omega}-\alpha]
\end{equation}
where the real function $G_{\alpha}^{I}(\tilde{q},\tilde{\omega})$ is described by
\begin{equation}\label{DYNPO_FUNCTION_IMAGINARY}
	G_{\alpha}^{I}(\tilde{q},\tilde{\omega}) \!=\! \frac{\left(\tilde{\omega}(\tilde{\omega}-2\alpha)-\tilde{q}^{2}\right)[(\tilde{\omega}-2\alpha)^{2}-\tilde{q}^{2}]}{\tilde{q}\sqrt{(\tilde{q}^{2}-\tilde{\omega}^{2})[(\tilde{\omega}-2\alpha)^{2}-\tilde{q}^{2}]}}
\end{equation}
and the coefficient $D_{\alpha}^{I}$ reads as follows:
\begin{equation}\label{DYNPO_COEFFICIENT_IMAGINARY}
	D_{\alpha}^{I}=-\alpha \; \Theta[(\tilde{q}^{2}-\tilde{\omega}^{2})\left\{(\tilde{\omega}-2\alpha)^{2}-\tilde{q}^{2}\right\}]
\end{equation}
Lastly, the coefficient $A$ is given by
\begin{equation}\label{DYNPO_PREFACTOR}
	A=\frac{1}{4} \, sgn[E_{F}] \, D(E_{F}) \, [\frac{k_{shift}}{k_{F}}] \, \cos{\theta_{\bm{q}}}\;.
\end{equation}
The $sgn[E_{F}]$ factor indicates that the drift-induced asymmetricity in the $E_{F} \! > \! 0$ ($E_{F} \! < \! 0$) case is solely dictated by the drift velocity, $\bm{v}_{dr}$, of the electron (hole) gas; hence, to have a better presentation of the physical aspects of this phenomenon, we rewrite the coefficient $A$ in terms of the drift velocity
\begin{equation}\label{DRIFT_VELOCITY}
	\bm{v}_{dr}\equiv \pm \frac{v_{F}}{\pi k_{F}^{2}} \int \Delta n_{F}[E^{\pm}(\bm{k})]\, \hat{\bm{k}} \, d^{2}\bm{k}\;.
\end{equation}
Plugging $\Delta n_{F}(E,\bm{k})$ from Eq. (\ref{FD_perturbation}), which corresponds to the case of $\hat{\bm{k}}_{shift} =-\hat{e}_{x}$, into Eq. (\ref{DRIFT_VELOCITY}) yields
\begin{equation}\label{GRPH_DRIFT_VELOCITY}
	\frac{\bm{v}_{dr}}{v_{F}}=sgn[E_{F}] \; \frac{\bm{k}_{shift}}{k_{F}}\;.
\end{equation}
Rewriting the coefficient $A$ in terms of the drift velocity from Eq. (\ref{GRPH_DRIFT_VELOCITY}) yields the prefactor appearing in Eq. (\ref{GRPH_CLOSED-FORM}). Fortunately, the expansive expressions for the real and imaginary parts of $\Delta\Pi(\bm{q},\omega)$ can be compacted into the complex function given by Eq. (\ref{GRPH_CLOSED-FORM}), which applies to complex frequency values. On the other hand, in each of the regions specified in Fig. \ref{REGIONS}, the real part of $\Delta\Pi(\bm{q},\omega)$ can be expressed in terms of real functions:
\begin{equation}\label{REAL_DYNPO_MODIFICATION_REGIONS}
	\Re{[\Delta\Pi \,]} \! \cong \! \frac{8A\tilde{\omega}}{\tilde{q}} \! + \! \frac{A\tilde{q}^{2}}{\sqrt{\left| \tilde{\omega}^{2} \! - \! \tilde{q}^{2} \right|}} \!\! \times \!\! \left\{
		\begin{aligned}
			0 \,\;\;\quad\quad &1 A
			\\
			H_{+} \! + \! H_{-} \quad &1 B
			\\
			-H_{+} \quad\quad & 2 A
			\\
			+H_{-} \quad\quad & 2 B
			\\
			H_{-} \! - \! H_{+}\quad & 3 A
			\\
			H_{-} \! - \! H_{+} \quad & 3 B
		\end{aligned}
	\right.
\end{equation}
Likewise, the imaginary part of $\Delta\Pi(\bm{q},\omega)$ is given by
\begin{equation}\label{IMAGINARY_DYNPO_MODIFICATION_REGIONS}
	\Im{[\Delta\Pi\,]} \! \cong \! -\frac{A \tilde{q}^{2}}{\sqrt{\left| \tilde{\omega}^{2} \! -\! \tilde{q}^{2} \right|}} \!\! \times \!\! \left\{
		\begin{aligned}
			H_{+}\! + \! H_{-}\quad & 1 A\quad\quad\;\;
			\\
			0 \,\;\;\quad\quad &1 B
			\\
			H_{-}\quad\quad & 2 A
			\\
			H_{+} \quad\quad & 2 B
			\\
			0 \,\;\;\quad\quad & 3 A
			\\
			0 \,\;\;\quad\quad &3 B
		\end{aligned}
	\right.\;.
\end{equation}
The real function $H_{\alpha}(\tilde{q},\tilde{\omega})$ is given by
\begin{equation}\label{H_ALPHA}
	H_{\alpha}(\tilde{q},\tilde{\omega}) \! = \! \left\{1 \! + \! \frac{\alpha\tilde{\omega}}{\tilde{q}} \! \left[ \! \frac{2 \! - \! \alpha\tilde{\omega}}{\tilde{q}} \! \right] \right\} \sqrt{\left|1 \! - \! \left[ \! \frac{2 \! -\! \alpha\tilde{\omega}}{\tilde{q}} \! \right]^{\! 2} \right|}\;.
\end{equation}
Having evaluated $\Delta\Pi$ in the $(\tilde{q},\tilde{\omega}>0)$ quarter-plane, we can evaluate $\Delta\Pi$ in the $(\tilde{q},-\tilde{\omega}<0)$ quarter-plane by exploiting the fact that the real and imaginary parts of $\Delta\Pi$ are odd and even functions of $\tilde{\omega}$, respectively. Note that this is just the opposite symmetry relation as compared to the equilibrium case.

\section{The Mermin's approach}\label{App:B}
Regarding the complexity of the electron scattering mechanisms, their impact on the response function can be approximately taken into account by replacing $\omega$ in the response function of the collisionless electron gas by $\omega+i\tau^{-1}$ with $\tau^{-1}$ being the phenomenological electron scattering rate which is related to the mobility of the graphene sample $\mu$ via the following relation: \cite{SARMA}
\begin{equation}\label{MOBILITY}
	\tau=\frac{e \, v_{F}^{2}}{\mu \left|E_{F}\right|}
\end{equation}
Such an imprecise scheme, however, fails to conserve the local electron number. The following correction formula removes such a defect for the case of the intrasubband longitudinal response function of the 2DEG \cite{Mermin}
\begin{equation}\label{MERMIN'S CORRECTION FORMULA}
	\Pi_{\tau}(\bm{q},\omega)=\frac{\Pi(\bm{q},\omega+i\tau^{-1})}{\displaystyle{1-\frac{1}{\displaystyle{1-i\omega\tau}}\left[1-\frac{\Pi(\bm{q},\omega+i\tau^{-1})}{\displaystyle{\Pi(\bm{q},0)}}\right]}}\;,
\end{equation}
where $\Pi(\bm{q},\omega)$ and $\Pi_{\tau}(\bm{q},\omega)$ denote the collisionless and the corrected dynamical polarization, respectively. Such a correction scheme has been shown to be applicable to the intraband dynamical polarization of graphene \cite{Hanson}; however, to our knowledge, there is no literature in which the application of Mermin's approach to the interband dynamical polarization of graphene is rigorously justified. Nonetheless, we relegate the clarification of this matter to the future works and follow the general trend of applying the Eq. (\ref{MERMIN'S CORRECTION FORMULA}) to the case of Dirac Fermions \cite{Marinko,Abajo,Koppens,Tanatar}. 

The dynamical polarization of the $\pi$ electron gas in a doped graphene sample at $T=0K$ is given by the following complex function: \cite{GAP_DYNPO}
\begin{equation}\label{ORG_DYNPOL}
	\Pi(q,\omega) \! = \! D(E_{F}) \left\{\frac{\tilde{q}^{2} \sum_{\alpha=\pm}G^{\alpha}(Z_{-\alpha})}{8 \sqrt{\tilde{q}^{2}-\tilde{\omega}^{2}}}-1\right\}\;,
\end{equation}
where $Z_{\alpha} \equiv (2-\alpha \tilde{\omega}^{\prime})/\tilde{q}$ and the complex function $G^{\alpha}(z)$ is defined as below:
\begin{equation}\label{G_{ALPHA}}
	G^{\alpha}(z) \equiv z\sqrt{1-z^{2}}+\alpha i \ln{[z+\sqrt{z-1}\sqrt{z+1}]}
\end{equation}
The effects of the disorder-induced electron scattering on the dynamical polarization of non-drifting $\pi$ electron gas in graphene can be taken into account by feeding the $\Pi(q,\omega\!+\!i\tau^{-1})$ values from Eq. (\ref{ORG_DYNPOL}) into Eq. (\ref{MERMIN'S CORRECTION FORMULA}). To obtain $\Pi_{\tau}^{\bullet}(\bm{q},\omega)$, we have computed $\Delta\Pi(\bm{q},\omega\!+\!i\tau^{-1})$ using Eq. (\ref{GRPH_CLOSED-FORM}), added it to $\Pi(q,\omega\!+\!i\tau^{-1})$ given by Eq. (\ref{ORG_DYNPOL}), and fed their sum into Eq. (\ref{MERMIN'S CORRECTION FORMULA}). The $\Pi_{\tau}^{\bullet}-\Pi_{\tau}^{\circ}$ values computed for a phenomenological scattering rate of $\hbar/\tau=5\,meV$ are presented in Fig. \ref{DP_MODIFICATION_PLOT}. For a Fermi energy of $100\,meV$, this $\tau$-value corresponds to a sample mobility of $\mu \approx 10^{4} \,\frac{cm^{2}}{V\cdot s}$, as suggested by Eq. (\ref{MOBILITY}). 

\section{The case of the 2DEG}\label{App:C}
Feeding the integral in Eq. (\ref{Dynpo}) with a single parabolic band, i.e. $E(\bm{k})\!=\!\hbar^{2}k^{2}/2m_{e}^{*}$ along with excluding $g_{v}$ and $f_{s,s'}\!\left(\bm{k},\bm{q}\right)$  yields the intrasubband dynamical polarization of the two-dimensional electron gas (2DEG) \cite{Stern}. Thanks to the absence of interband transitions and the parabolic energy dispersion, a change of the integration variable according to $\bm{k}'= \bm{k} - \bm{k}_{shift}$ yields an explicit relation between $\Pi^{\bullet}(\bm{q},\omega)$ and $\Pi^{\circ}(\bm{q},\omega)$ given by Eq. (\ref{2DEG_DYNPO_MOD_EXACT}). In spite of the fact that the transformation given by Eq. (\ref{2DEG_DYNPO_MOD_EXACT}) provides the best possible approximation within the framework of the shifted Fermi disk model, we present the 2DEG counterpart of the expression given by Eq. (\ref{GRPH_CLOSED-FORM}) in order to have a comparative picture:
\begin{equation}\label{2DEG_CLOSED-FORM}
	\Delta\Pi(\bm{q},\omega) \! \cong \! \frac{\bm{q} \cdot \bm{v}_{dr}}{q \, v_{F}} \, \frac{D(E_{F})}{\tilde{q}} \! \sum_{\alpha=\pm}\!\!\frac{\alpha}{\sqrt{\displaystyle{1 \! - \! \left[\frac{2\tilde{q}}{\tilde{\omega}^{\prime}-\alpha\tilde{q}^{2}}\right]^{2}}}}	
\end{equation}
with $D(E_{F})\!=\! g_{s} m_{e}^{*}/2\pi \hbar^{2}$ being the DOS of the parabolic band and $\bm{v}_{dr}= v_{F} [\bm{k}_{shift}/k_{F}]$ is the drift velocity.

\section{The onset of the Landau damping in the absence of drift}\label{App:D}
In the absence of drift, the TM-SPP dispersion can be implicitly obtained via plugging the analytic expression for the dynamical polarization given by Eq. (\ref{ORG_DYNPOL}) into Eq. (\ref{TM-SPP_DISPERSION}). The onset of the Landau damping $q_{c}$ is the $q$-value where the TM-SPP dispersion curve and the line $\tilde{\omega}=2-\tilde{q}$ meet which is yielded by the below equation:
\begin{equation}\label{q_c}
	\frac{1}{16} \frac{\tilde{q}_{c}^{2}}{\sqrt{1-\tilde{q}_{c}}} \; G_{>}\left(\frac{4}{\tilde{q}_{c}}-1\right)=1+\frac{\tilde{q}_{c}(\epsilon_{1}+\epsilon_{2})}{2 g_{s} g_{v} \alpha_{f} (c/v_{F})}
\end{equation}
where the function $G_{>}(x)$ is given by
\begin{equation}\label{G_>(x)}
	G_{>}(x)=x\sqrt{x^{2}-1}-\ln{[x+\sqrt{x^{2}-1}]}\;.
\end{equation}
\bibliography{DRAFT_BIB}
\end{document}